\batchmode
\documentclass[twocolumn,showpacs,preprintnumbers,amsmath,amssymb]{revtex4}

\usepackage{graphicx}
\usepackage{dcolumn}
\usepackage{bm}

\begin{document}

\preprint{APS/123-QED}


\title{Coherent description of the intrinsic and extrinsic
anomalous Hall effect in disordered alloys
on an $ab$ $initio$ level}

\author{S. Lowitzer}
\author{D. K\"odderitzsch}
\author{H. Ebert}
\affiliation{%
Department Chemie, Physikalische Chemie, Universit\"at M\"unchen, Butenandtstr. 5-13, 81377 M\"unchen, Germany\\
}%

\date{\today}

\begin{abstract}
A coherent description of the anomalous Hall effect (AHE) is presented
that is applicable to pure as well as disordered alloy systems by
treating all sources of the AHE on equal footing. This is achieved by
an implementation of the Kubo-St\v{r}eda equation using the fully
relativistic Korringa-Kohn-Rostoker (KKR) Green's function method in
combination with the Coherent Potential Approximation (CPA) alloy
theory. Applications to the pure elemental ferromagnets bcc-Fe and
fcc-Ni led to results in full accordance with previous work. For the
alloy systems fcc-Fe$_x$Pd$_{1-x}$ and fcc-Ni$_x$Pd$_{1-x}$ very
satisfying agreement with experiment could be achieved for the
anomalous Hall conductivity (AHC) over the whole range of
concentration. To interpret these results an extension of the
definition for the intrinsic AHC is suggested. Plotting the
corresponding extrinsic AHC versus the longitudinal conductivity a
linear relation is found in the dilute regimes, that allows a detailed
discussion of the role of the skew and side-jump scattering processes.
\end{abstract}

\maketitle
During the last years the anomalous Hall effect (AHE) has received great
 interest. This is partly caused by its close connection to the spin
 Hall effect (SHE), that possesses a large potential for application in
 the rapidly growing field of spintronics
\cite{AF07}. On the other hand, many theoretical investigations
 are devoted to the development of a coherent description of these
quite complex phenomena \cite{NSO+10}.

As was already pointed out by Karplus and Luttinger \cite{KL54} the
ultimate origin for the AHE in ferromagnets is the spin-orbit coupling
(SOC) that -- together with the spontaneous magnetization -- leads to
a symmetry breaking. As was demonstrated by experiment
\cite{Dhe67,Lav61} and is obvious from the work of Karplus and
Luttinger the AHE is present even in pure systems. This so-called
intrinsic AHE could later be connected to the Berry-phase
\cite{FNT+03} and corresponding $ab$ $initio$ results could be obtained
during the last years using an expression for the anomalous Hall
conductivity (AHC) $\sigma_{\rm xy}$ in terms of the Berry curvature
\cite{YKM04,WVYS07}.  For diluted and concentrated alloys, on the
other hand, the occurrence of the AHE was primarily ascribed to the
spin-dependent skew or Mott \cite{Smi55,Smi58} and the so-called
side-jump \cite{Ber70} scattering mechanisms. The latter one is caused
by the anomalous velocity, a first-order relativistic correction to
the non-relativistic velocity operator connected to
SOC. Interestingly, scaling laws connecting the AHC $\sigma_{\rm xy}$
and the longitudinal conductivity $\sigma_{\rm xx}$ (see below) could
be derived for these two extrinsic mechanisms \cite{NSO+10}. Their
treatment in connection with a description of electronic transport in
terms of wave packet dynamics was discussed in detail recently by
Sinitsyn \cite{Sin08}.
When dealing with the extrinsic AHE in disordered systems, however,
disorder was treated so far only by model potentials \cite{KST10} or by
a damping parameter \cite{OSN06,OSN08}.\\ Cr\'epieux and Bruno \cite{CB01}
performed qualitative investigations on the AHE on the basis of the
Kubo-St\v{r}eda equation.  This equation is derived from Kubo's linear
response formalism supplying a suitable basis for investigations based
on a realistic description of the underlying electronic structure (see
Ref.\ \onlinecite{NHK10} and below). An alternative description of the
AHE with a wider regime of applicability is achieved by using the
non-equilibrium Green's function formalism.  Using a suitable, but
still tractable, model description for the electronic structure Onoda
et al.\ \cite{OSN06,OSN08} could divide the range of $\sigma_{\rm xx}$
covered typically by real materials into three regimes with different
scaling laws connecting $\sigma_{\rm xy}$ and $\sigma_{\rm xx}$.

In this communication results for the AHC obtained using the
 Kubo-St\v{r}eda equation are presented.  Using a fully relativistic
 Green's function formulation in combination with a reliable alloy
 theory a coherent description for pure as well as diluted and
 concentrated alloys could be achieved that treats intrinsic and
 extrinsic sources of the AHE on equal footing.

The Kubo linear response formalism supplies an appropriate basis to
 deal with electronic transport in magnetic metallic systems.  Making
 use of a single-particle description of the electronic structure and
 restricting to the case $T=0$~K one is led to the
 Kubo-St\v{r}eda equation
  for the electrical conductivity tensor ${\bm
 \sigma}$ \cite{Str82}. For cubic systems with the magnetization along
 the $z$-direction, the AHE is described by the corresponding
 off-diagonal tensor element or anomalous Hall conductivity 
 $\sigma_{\rm xy}$ given by:\cite{Str82,CB01a}
\begin{eqnarray}
\label{eq:streda}
\sigma_{\rm xy}& =& \frac{\hbar }{4\pi N\Omega}
                 {\rm Trace}\,\big\langle \hat{j}_{\rm x} (G^+-G^-) \hat{j}_{\rm y}  G^-  
                      \nonumber   \\  
&&  \qquad \qquad \quad 
-  \hat{j}_{\rm x} G^+\hat{j}_{\rm y}(G^+-G^-)\big\rangle_{\rm c} \nonumber \\
&& + \frac{e}{4\pi i N\Omega} {\rm Trace}\,
\big\langle (G^+-G^-)(\hat{r}_{\rm x}\hat{j}_{\rm y} - \hat{r}_{\rm y}\hat{j}_{\rm x}) \big\rangle_{\rm c} 
\label{eq:bru} 
\; .
\end{eqnarray}
 Here $\Omega$ is the volume of the unit cell, $N$ is the number of
 sites, while $ \hat{\bf r} $ and $\hat{\bf j}$ are the position and
 current density operators, respectively.
 For the cubic systems considered here the last term is site-diagonal
 for symmetry reasons. As furthermore
all system considered here are metallic it has been
omitted \cite{NHK10}.
 The electronic structure of
 the system is represented in terms of the single-particle retarded
 ($G^+$) and advanced ($G^-$) Green's functions at the Fermi energy
 $E_{\rm F}$.
 Within the present work these functions have been
 evaluated by means of the multiple scattering Korringa-Kohn-Rostoker
 (KKR) formalism \cite{Ebe00}. The chemical disorder in the
 investigated random substitutional alloys has been accounted for by
 using the Coherent Potential Approximation (CPA) \cite{Sov67}.  This
 alloy theory supplies a reliable framework to perform the
 configurational average indicated by the brackets $\langle
 ... \rangle_{\rm c} $ in Eq.~(\ref{eq:bru}).  It includes, in
 particular, a clear definition for differences of configurational
 averages like $ \langle \hat{j}_{\rm x} G^+ \hat{j}_{\rm y} G^-
 \rangle_{\rm c} - \langle \hat{j}_{\rm x} G^+ \rangle_{\rm c} \langle
 \hat{j}_{\rm y} G^- \rangle_{\rm c} $.  These so-called
 vertex-corrections correspond to the scattering-in terms within
 semi-classical Boltzmann transport theory \cite{But85}.

 Dealing with the AHE requires to account for the influence of
 spin-orbit coupling in an appropriate way. This is achieved by using
 the four-component Dirac formalism.\cite{Ros61} In combination with
 spin-density functional theory in its local approximation (LSDA) the
 corresponding Dirac Hamiltonian is given by:\cite{MV79}
\begin{equation}
\label{eq:Dirac}
{\cal H}_{\rm D} = c {\bm \alpha}\cdot\hat{\bf p}
 + \beta mc^2 +V +\beta \Sigma_z B
 \; .
\end{equation}
 Here $\hat{\bf p}=-i\hbar{\bm \nabla}$ is the canonical momentum
operator, ${\bm \alpha}$ and $\beta $ are the standard Dirac
matrices,\cite{Ros61} while $V$ and $B$ represent the spin-independent
and spin-dependent, respectively, effective LSDA potentials for the
magnetization along $z$.  Within the fully relativistic framework
adopted here the current density operator $\hat{\bf j}$ is given
by:\cite{Ros61}
\begin{equation}
\label{eq:j}
\hat{\bf j}  = -  c \, | e |\,  {\bm \alpha} \; .
\end{equation}
To allow for a more detailed discussion on the origin of the AHE it is
useful to introduce the alternative current density
operator:\cite{GE95}
\begin{eqnarray}
\label{eq:reform}
\hat{\bf j}_{ \hat{\bf p} }   & =&  
\frac{  - | e | }{m+E/c^2}    \bigg( 
 \hat{\bf p} + \frac{V}{c} {\bm \alpha} 
+\frac{B}{c}\beta\Sigma_z(\alpha_x,\alpha_y,0)^{\rm T}   \bigg)
\;,
\end{eqnarray}
that is equivalent to $\hat{\bf j}$ given by Eq.~(\ref{eq:j}) and that
can be derived from the anti-commutator of $\hat{\bf j}$ and the Dirac
Hamiltonian ${\cal H}_{\rm D}$ given by Eq.~(\ref{eq:Dirac}).

\medskip

Recently, the intrinsic AHE of the pure ferromagnets Fe, Co and Ni
 \cite{YKM04,WVYS07,RMS09} as well as ordered FePt and FePd \cite{SMA+10} has
 been investigated theoretically on an $ab$ $initio$ level using the
 formulation for $\sigma_{\rm xy}$ in terms of the Berry
 curvature. Alternatively, the tensor element $\sigma_{\rm xy}$ can be
 obtained directly from the expression given in Eq.~(\ref{eq:bru})
 that is evaluated by Fourier transformation leading to a
 corresponding Brillouin zone integration \cite{But85}. As the
 integrand shows a $\delta$-function like behavior for pure systems a
 small imaginary part $\epsilon$ has to be added to the Fermi energy
 $E_{\rm F}$ and an extrapolation to zero has to be made for
 $\epsilon$.  For the calculations of $\sigma_{\rm xy}$ performed for
 bcc-Fe and fcc-Ni $\epsilon$ has been varied between $10^{-3}$ and
 $10^{-6}$ Ry. To ensure convergence of the Brillouin zone integration
 about $10^9$ ${\bf k}$-points have been used.  The resulting AHC of
 bcc-Fe and fcc-Ni is given in Table I together with experimental data
 as well as results of previous $ab$ $initio$
 work.\cite{YKM04,WVYS07,xxx}
%
\begin{table}[]
\centering
\caption{\label{tab:intr}The intrinsic AHC of bcc-Fe and fcc-Ni from $ab$
 $initio$ theoretical as well as experimental (Exp.) investigations. }
\begin{tabular}{lcccccc}
\hline
\hline
$\sigma_{\rm xy}$ $({\rm m}\Omega \, {\rm cm})^{-1}$ &&&& bcc Fe     && fcc Ni \\
\hline
present work &\phantom{aaaaaaaaaaaaaaaaa}  &&&  0.638 &\phantom{aaaa}&-1.635 \\
\citet{YKM04}   &&&& 0.753 && \\
\citet{WVYS07}  &&&& 0.751 && -2.203 \\
Yao~\cite{xxx}     &&&&       && -2.073\\
Exp. \cite{Dhe67,Lav61}      &&&& 1.032 && -0.646 \\
\hline 
\hline
\end{tabular}
\end{table}
%
Taking into account that SOC was treated by approximate schemes
within the  corresponding  calculations, the agreement  is quite satisfying.

The expression for $\sigma_{\rm xy}$ in terms of the Berry curvature
used within previous work is completely equivalent to the
Kubo-St\v{r}eda equation used here, as both approaches are based on
Kubo's linear response formalism and adopt a single-particle
description for the electronic structure for $T=0$~K.\cite{NHK10}
However, it should be stressed that the Berry curvature is usually
formulated in terms of Bloch states implying translational symmetry
this way. Calculations of $\sigma_{\rm xy}$ for disordered alloys with
broken translational symmetry are therefore not possible on this basis
while the Kubo-St\v{r}eda equation supplies an adequate framework for
such investigations.  As the ${\bf k}$-dependent integrand connected
with Eq.~(\ref{eq:bru}) gets smeared out in this case due to the
chemical disorder it looses its $\delta$-function like behavior.  For
that reason, broadening via a complex Fermi energy with $\epsilon>0$
is not necessary for calculations on alloys.

\begin{figure}
 \begin{center}
 \includegraphics[width=8cm,clip]{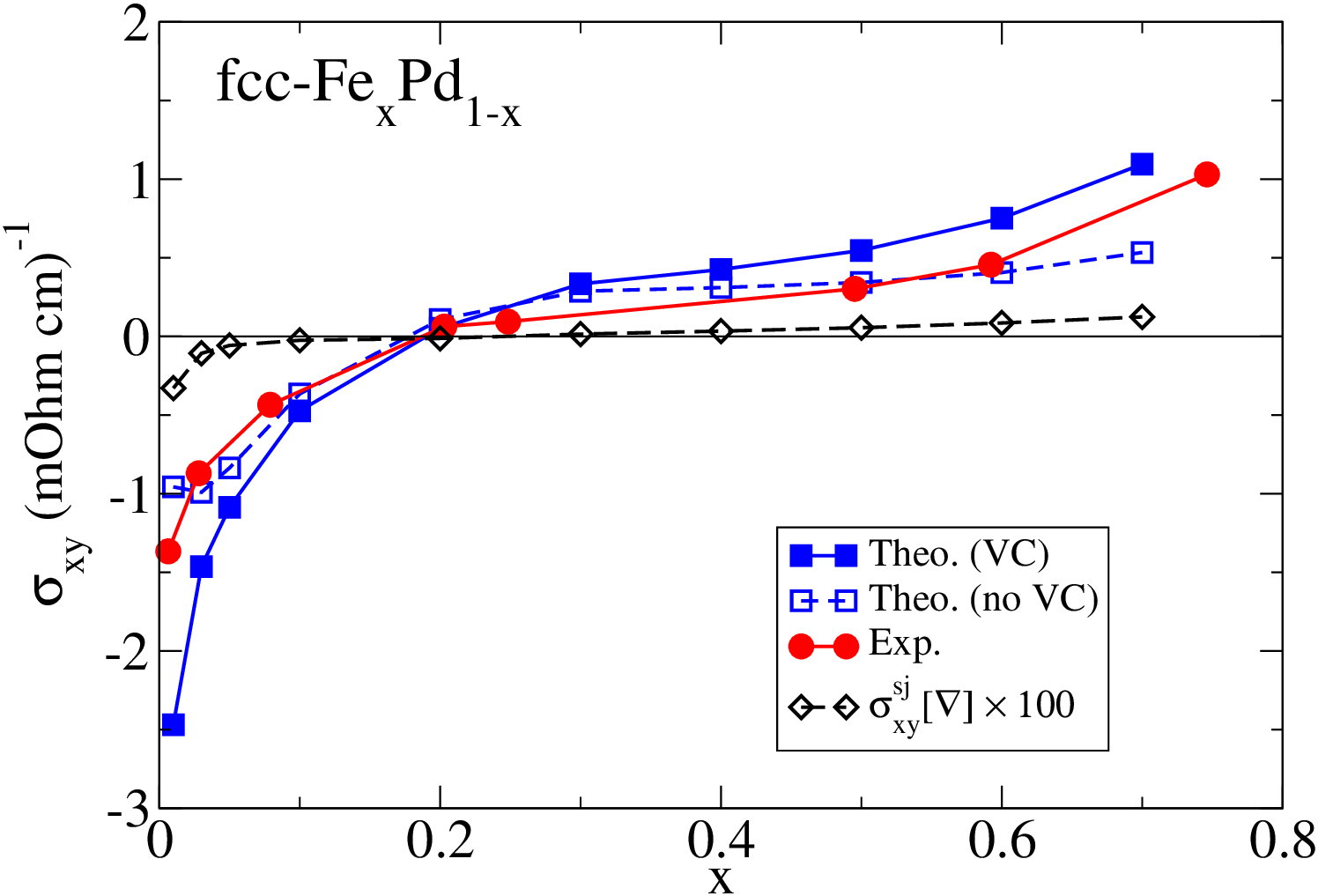}
 \vspace*{0.3cm}

 \includegraphics[width=8cm,clip]{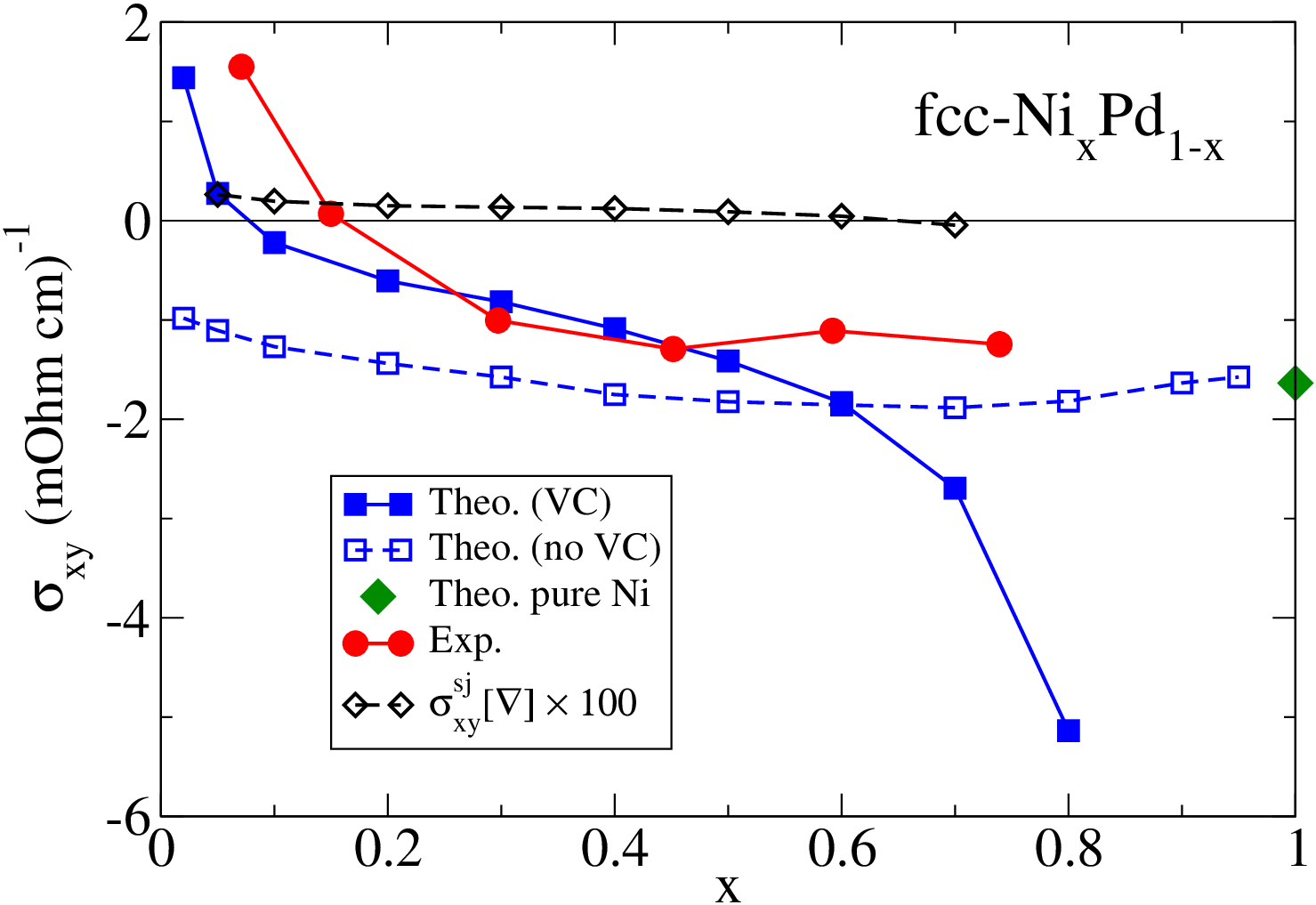} \caption{\label{plot:AHE}(Color online) 
The AHC  of fcc-Fe$_x$Pd$_{1-x}$ and fcc-Ni$_x$Pd$_{1-x}$. 
The  total AHC $\sigma_{\rm xy}$  (full squares) has been 
calculated including the vertex corrections 
($\sigma_{\rm xy} \equiv \sigma_{\rm xy}^{\rm VC}$),
while the intrinsic AHC $\sigma_{\rm xy}^{\rm intr} $ (open squares)
has been obtained by omitting them 
($\sigma_{\rm xy}^{\rm intr} \equiv \sigma_{\rm xy}^{\rm no\, VC}$).
 In addition, 
experimental data \cite{MF82}  for $\sigma_{\rm xy}$  (full circles)
determined  at $T=4.2$~K are shown.
 Further, an estimation for the side-jump
 contribution $ \sigma_{\rm xy}^{\rm sj}$ to the extrinsic AHC of
 fcc-Fe$_x$Pd$_{1-x}$ and fcc-Ni$_x$Pd$_{1-x}$ calculated as the
 difference $ \sigma_{\rm xy}^{\rm extr} - \sigma_{\rm xy}^{\rm
 extr}[\nabla] $ is shown (open diamonds; see text).
}
 \end{center}
\end{figure}

Within the present work, corresponding calculations have been done
for the alloy systems fcc-Fe$_x$Pd$_{1-x}$ and fcc-Ni$_x$Pd$_{1-x}$.
Both systems are formed by the nearly ferromagnetic
transition metal Pd with an elemental 3d-ferromagnet
leading to 
 a very low critical concentration   $x_{\rm crit}$ for the on-set
of spontaneous ferromagnetic order 
($x_{\rm crit}^{\rm  Fe}\approx 0$ for   Fe$_x$Pd$_{1-x}$
  and $x_{\rm crit}^{\rm Ni}\approx 0.02$ for   Ni$_x$Pd$_{1-x}$).
As can be seen in Fig.~\ref{plot:AHE}, calculation of the AHC via
 Eq.~(\ref{eq:streda}) leads to a very satisfying agreement with the
experimental data \cite{MF82}, in particular, for Fe$_x$Pd$_{1-x}$,
 that are available over a wide range of composition.
%
%
In particular the change in sign of $\sigma_{\rm xy}$ with composition
observed for both alloy systems is well reproduced by the
calculations.  As one might speculate from the data for elemental
bcc-Fe and fcc-Ni in Table I the sign of $\sigma_{\rm xy}$ on the
Pd-poor side of fcc-Fe$_x$Pd$_{1-x}$ is indeed positive while it is
negative for fcc-Ni$_x$Pd$_{1-x}$.  For the Pd-rich the situation is
reversed, clearly showing that the elemental ferromagnet is the
primary source for the AHE in these two alloy systems (see below).

To get a more detailed insight into the mechanism responsible for the
AHE in the investigated alloys, a decomposition of the AHC has been
performed. A formal basis for this is provided by the representation
of the Kubo-St\v{r}eda equation in terms of Feynman
diagrams.\cite{CB01} From this it can be seen that the skew and
side-jump mechanisms are exclusively connected to diagrams involving
the vertex corrections. The remaining diagrams are standing for
products of the type $ {\langle\hat{j}_{\rm x}G^+\rangle_{\rm c}}
{\langle\hat{j}_{\rm y}G^-\rangle_{\rm c}} $ that correspond to the
intrinsic AHE and correction terms due to chemical disorder. It seems
therefore sensible to extend the definition of the intrinsic AHC
$\sigma_{\rm xy}^{\rm intr}$ to the case of diluted and concentrated
alloys by combining all contributions not connected to the vertex
corrections.  This obviously allows to calculate the total AHC
($\sigma_{\rm xy}$) and the intrinsic one ($\sigma_{\rm xy}^{\rm
intr}$) by evaluating the Kubo-St\v{r}eda equation
(Eq.~(\ref{eq:bru})) with and without, resp., including the vertex
corrections (VC); i.e.\ identifying $\sigma_{\rm xy} \equiv
\sigma_{\rm xy}^{\rm VC}$ and $\sigma_{\rm xy}^{\rm intr} \equiv
\sigma_{\rm xy}^{\rm no\, VC}$, respectively. 
As seen in Fig.~\ref{plot:AHE}, $\sigma_{\rm xy}^{\rm intr}$ gives a
major contribution to the total AHC $\sigma_{\rm xy}$ of
fcc-Fe$_x$Pd$_{1-x}$ and shows in particular also a change in sign
with varying concentration.  For fcc-Ni$_x$Pd$_{1-x}$, on the other
hand, $\sigma_{\rm xy}^{\rm intr}$ varies weakly with composition and
extrapolates rather well to the intrinsic AHC of pure Ni (see Table
I). For both alloy systems $\sigma_{\rm xy}^{\rm intr} \approx 1\,({\rm m}\Omega {\rm cm})^{-1}$ when $x_{\rm Pd}$ approaches 1
indicating that the intrinsic AHC is primarily determined by the
properties of the Pd-host in the dilute regime. These findings
obviously justify the extension of the definition for $\sigma_{\rm
xy}^{\rm intr}$ to represent all contributions not connected to the
vertex corrections.

The longitudinal conductivity $\sigma_{\rm xx}$ of
 fcc-Fe$_x$Pd$_{1-x}$ and fcc-Ni$_x$Pd$_{1-x}$ lies nearly exclusively
 in the so-called super-clean regime with $\sigma_{\rm xx} \gtrsim
 (\mu\Omega {\rm cm})^{-1}$.\cite{OSN06,OSN08} For this regime the
 skew scattering mechanism should dominate $\sigma_{\rm xy}$ obeying
 the relation $ \sigma_{\rm xy} = S \, \sigma_{\rm xx} $, with $S$
 being the so-called skewness factor \cite{OSN06,OSN08}. Accounting
 for all three mechanisms one is therefore led to the
 decomposition:\cite{NSO+10}
%
\begin{equation}
\label{eq:xy}
\sigma_{\rm xy}= 
\sigma_{\rm xy}^{\rm intr}+ 
 S \,\sigma_{\rm xx} + \sigma_{\rm xy}^{\rm sj}
=  \sigma_{\rm xy}^{\rm intr}+ \sigma_{\rm xy}^{\rm extr}
 \; ,
\end{equation}
%
that may be seen as a definition for the side-jump contribution
$\sigma_{\rm xy}^{\rm sj}$.\cite{NSO+10}
In fact, a plot of $\sigma_{\rm xy}$ versus $\sigma_{\rm xx}$
with the concentration as an implicit parameter
was used in the past to decompose the experimental AHC
of alloy systems accordingly.\cite{MB73,SOT09,TYJ09}

\begin{figure}
 \begin{center}
 \includegraphics[height=5.5cm,clip]{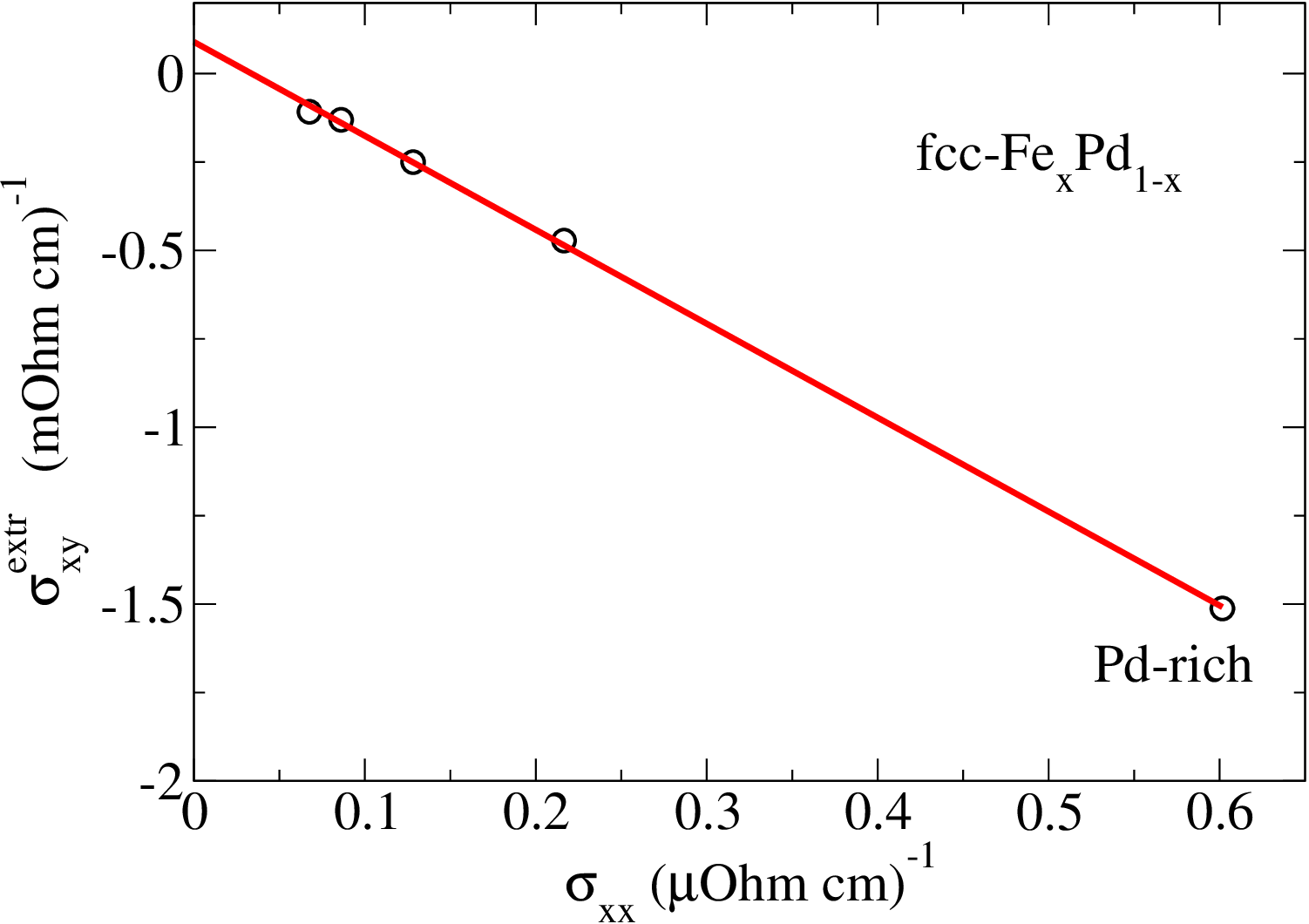}
 \vspace*{0.3cm}

 \includegraphics[height=5.5cm,clip]{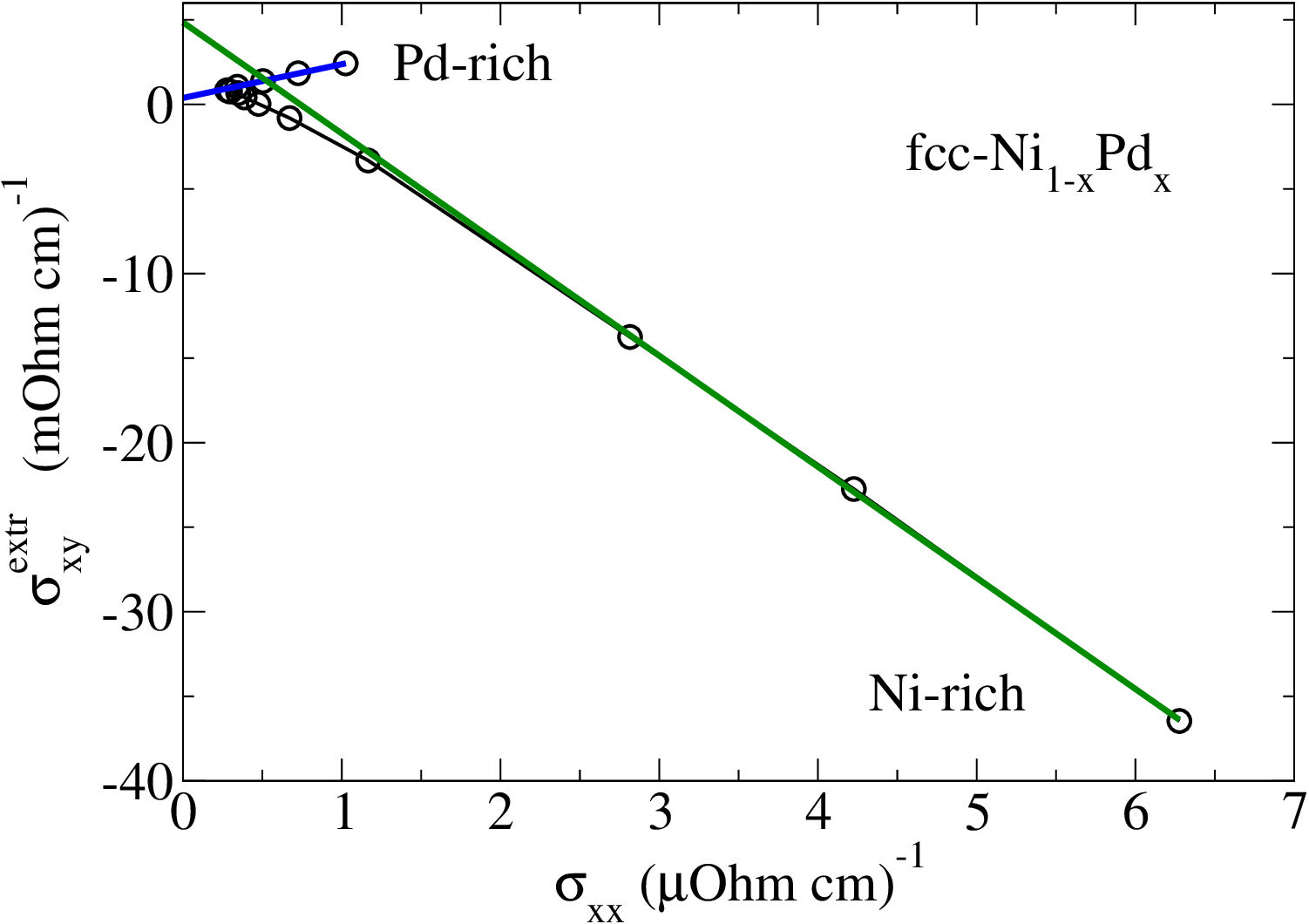}
 \caption{\label{plot:extr}(Color online) The extrinsic AHC
 $\sigma_{\rm xy}^{\rm extr}$ versus $\sigma_{\rm xx}$ for
 fcc-Fe$_x$Pd$_{1-x}$ and fcc-Ni$_x$Pd$_{1-x}$.  The straight lines
 represent extrapolations of the data for $x_{\rm Pd} \geq 0.9$ (Pd-rich)
 and $x_{\rm Ni} \geq 0.9$ (Ni-rich) to $\sigma_{\rm xx}=0$.}
 \end{center}
\end{figure}

In Fig.~\ref{plot:extr} the extrinsic AHC of fcc-Fe$_x$Pd$_{1-x}$ and
fcc-Ni$_x$Pd$_{1-x}$ defined as $\sigma_{\rm xy}^{\rm
extr}=\sigma_{\rm xy}- \sigma_{\rm xy}^{\rm intr}$ is plotted versus
the longitudinal conductivity $\sigma_{\rm xx}$.
%
%
 Obviously, the relation suggested by Eq.~(\ref{eq:xy}) is well
fulfilled on the Pd-rich side of both systems as well as on the
Ni-rich side of Ni$_x$Pd$_{1-x}$. Extrapolating for these regimes to
$\sigma_{\rm xx}=0$ allows to deduce the corresponding skewness
parameters and side-jump term $ \sigma_{\rm xy}^{\rm sj}$
(Fe$_x$Pd$_{1-x}$ for $x_{\rm Pd}\geq 0.9$: $S=-2.7\cdot 10^{-3}$ and $ \sigma_{\rm
xy}^{\rm sj} \approx 0.1\,({\rm m}\Omega {\rm cm})^{-1} $;
Ni$_x$Pd$_{1-x}$ for $x_{\rm Pd}\geq 0.9$: $S=2.0\cdot 10^{-3}$ and $
\sigma_{\rm xy}^{\rm sj} \approx 0.4\,({\rm
m}\Omega {\rm cm})^{-1} $ and for $x_{\rm Ni}\geq 0.9$: $S=-6.6\cdot 10^{-3}$ and $
\sigma_{\rm xy}^{\rm sj} \approx 4.8\,({\rm m}\Omega {\rm cm})^{-1} $).
These results show clearly that the skew scattering mechanisms by far
dominates $\sigma_{\rm xy}^{\rm extr}$ in the dilute regimes.  For the
two alloy systems the corresponding skewness factor $S$ is found
comparable in magnitude but different in sign on the Pd-rich side (see
above). This once more demonstrates that the skew scattering mechanism
has to be associated primarily with the solute component Fe or Ni,
respectively.

As emphasized above, Eq.~(\ref{eq:xy}) can be seen as a definition for 
various extrinsic contributions to $\sigma_{\rm xy}^{\rm extr}$
according to their scaling behavior.
An alternative way to define the side-jump term $ \sigma_{\rm xy}^{\rm sj}$
is to make use of its connection with the anomalous velocity,
that is  a correction to the non-relativistic
current density operator
$\hat{\bf j}_{\rm nr} = - \frac{| e |}{m} \frac{\hbar}{i} {\bm \nabla}$.
Within the relativistic approach used here, an estimate 
for $ \sigma_{\rm xy}^{\rm sj}$ can be made using the 
 alternative current density   operator
  $ \hat{\bf j}_{ \hat{\bf p} } $
with  the potential 
terms $V$ and $B$ suppressed (see Eq.~(\ref{eq:reform})). 
The corresponding extrinsic AHC
 $\sigma_{\rm xy}^{\rm extr}[\nabla]  =  \sigma_{\rm xy}[\nabla]  
-  \sigma_{\rm xy}^{\rm intr}[\nabla]$ allows to write
$ \sigma_{\rm xy}^{\rm sj}\approx \sigma_{\rm xy}^{\rm sj}[\nabla] =  
    \sigma_{\rm xy}^{\rm extr} -  \sigma_{\rm xy}^{\rm extr}[\nabla]      $.
The results for  $ \sigma_{\rm xy}^{\rm sj}$ obtained this way for 
fcc-Fe$_x$Pd$_{1-x}$ and fcc-Ni$_x$Pd$_{1-x}$
are also shown in Fig.~\ref{plot:AHE}.
As one notes there is obviously a non-negligible concentration
dependency for both alloy systems in particular on the Pd-rich
side. In both cases, however, the numerical results are much smaller
than for the intrinsic as well as the skew-scattering contributions.
While this once more supports the conclusion that the AHE of the
investigated alloy systems is dominated by the latter mechanisms, it
also shows that the quantitative results for the side-jump term $
\sigma_{\rm xy}^{\rm sj}$ may depend strongly on the definition used.

In summary, a coherent description of 
 the AHE for pure metals and
diluted as well as  concentrated alloys on an 
$ab$ $initio$ level was presented  based on a
 fully relativistic implementation of the 
Kubo-St\v{r}eda equation
 using the multiple-scattering
or KKR formalism in combination with the CPA alloy theory.
The intrinsic AHC obtained this way for bcc-Fe and fcc-Ni
was found in satisfying agreement with previous $ab$ $initio$ 
work using an equivalent expression for $ \sigma_{\rm xy}$ 
in terms of the Berry curvature. Corresponding calculations 
for the alloy systems fcc-Fe$_x$Pd$_{1-x}$ and fcc-Ni$_x$Pd$_{1-x}$
reproduced the available experimental data very well.
Identifying the contributions to  $ \sigma_{\rm xy}$ 
that are not connected to the vertex corrections with the
intrinsic AHE of an alloy allowed to decompose the remaining
extrinsic AHE. Plotting   $ \sigma_{\rm xy}$ 
versus   $ \sigma_{\rm xx}$  
it was found that the skew scattering
term by far dominates the side-jump
contribution in the dilute alloy regime. 
This conclusion could be supported by model calculations
that supplied an estimate for the contribution
to   $\sigma_{\rm xy}$  due to the anomalous velocity.

\medskip

{\bf Acknowledgments}

The authors would like to thank the DFG for financial support within
the SFB 689 ``Spinph\"anomene in reduzierten Dimensionen'' for
financial support.

\end{document}